\def\year{2021}\relax
\documentclass[letterpaper]{article} 
\usepackage{aaai21}  
\usepackage{times}  
\usepackage{helvet} 
\usepackage{courier}  
\usepackage[hyphens]{url}  
\usepackage{graphicx} 
\usepackage{natbib}  
\usepackage{caption} 
\usepackage{amsmath}
\usepackage{booktabs}
\usepackage{algorithm}
\usepackage{algorithmic}
\usepackage[utf8]{inputenc} 
\usepackage[T1]{fontenc} 
\usepackage{amsfonts} 
\usepackage{stmaryrd}
\usepackage{subfigure}
\usepackage[switch]{lineno}  %

\renewcommand{\vec}[1]{\ensuremath{\mathbf{#1}}}

\newcommand{\mat}[1]{\ensuremath{\mathbf{#1}}}

\newcommand{\ten}[1]{\mat{\ensuremath{\boldsymbol{\mathcal{#1}}}}}

\newcommand{\norm}[1]{\left\lVert#1\right\rVert}
\newcommand{\outprod}{\circ}

\title{PANTHER: Pathway Augmented Nonnegative Tensor factorization for HighER-order feature learning}
\author{Yuan Luo\thanks{Corresponding Author}, Chengsheng Mao\\
}
\affiliations{
Northwestern University\\
Chicago IL 60611\\
\{yuan.luo, chengsheng.mao\}@northwestern.edu\\
}

\begin{document}
\maketitle

\begin{abstract}
Genetic pathways usually encode molecular mechanisms that can inform targeted interventions. It is often challenging for existing machine learning approaches to jointly model genetic pathways (higher-order features) and variants (atomic features), and present to clinicians interpretable models. In order to build more accurate and better interpretable machine learning models for genetic medicine, we introduce Pathway Augmented Nonnegative Tensor factorization for HighER-order feature learning (PANTHER). PANTHER selects informative genetic pathways that directly encode molecular mechanisms. We apply genetically motivated constrained tensor factorization to group pathways in a way that reflects molecular mechanism interactions. We then train a softmax classifier for disease types using the identified pathway groups. We evaluated PANTHER against multiple state-of-the-art constrained tensor/matrix factorization models, as well as group guided and Bayesian hierarchical models. PANTHER outperforms all state-of-the-art comparison models significantly ($p<0.05$). Our experiments on large scale Next Generation Sequencing (NGS) and whole-genome genotyping datasets also demonstrated wide applicability of PANTHER. We performed feature analysis in predicting disease types, which suggested insights and benefits of the identified pathway groups.
\end{abstract}

\section{Introduction}
Genetic medicine aims to use similar patients' genetic profiles to infer the disease status or the treatment options, thus it is critical to accurately define patient similarity. Such similarity can be captured not only by the atomic features (e.g., genetic variants), but also by the higher-order features (e.g., genetic pathways) describing the relationships among the atomic features. Intuitively, the genetic pathways interact with each other and collectively drive pathogenesis, but their knowledge largely relies on curation and is still evolving. In contrast, genetic variants are routinely analyzed for robust statistical associations with diseases~\cite{ellrott2018scalable}; however, they usually do not themselves correspond to molecular mechanisms, which often leads to lack of functional interpretability. 

Genetic pathways are a valuable tool to assist in representing, understanding, and analyzing the complex interactions between molecular mechanisms. The pathways encode genetic functions including regulations, genetic signaling, and metabolic interactions. They have a wide range of applications including predicting cellular activity and inferring disease types and status. Individual pathways are themselves part of the entire biological system and interact with each other. For example, a signaling pathway sensing the environment may govern the expression of transcription factors in another signaling pathway, which then controls the expression of proteins that play roles as enzymes in a metabolic pathway. So it is important to model co-functioning molecular mechanisms~\cite{alon2019introduction}. Grouping pathways together as features provides insights on interacting molecular mechanisms. Atomic features can potentially help to better correlate higher-order features (e.g., common genes help correlate pathways). Thus joint consideration of higher-order features and atomic features is beneficial.

Clinicians often regard existing machine learning models for genetics as hard-to-interpret black boxes. Existing machine learning approaches in genetics usually do not group interacting genetic pathways together as features. In fact, few machine learning algorithms jointly model higher-order features and atomic features, and most algorithms adopt a flat \textit{subject} $\times$ \textit{feature} matrix view.  It is already challenging to automate the exploration of interactions between atomic-features (for which a series of matrix and tensor factorization methods were proposed~\cite{wang2012nonnegative,kolda2009tensor}), not to mention the exploration of interactions between higher-order features. Although theoretically one can add interactions as additional features or embed graphical models to account for feature interactions, the problem quickly becomes intractable for multiple feature modalities (e.g., higher-order and atomic features) and large feature dimensionality (e.g. at the genome scale).

In this paper, we propose a novel framework named PANTHER: a Pathway Augmented Nonnegative Tensor factorization for HighER-order feature learning to group interacting higher-order genetic pathways as features. To promote reproducibility, we share our source code at https://github.com/yuanluo/panther. Contributions of this paper are:
\begin{itemize}
\setlength\itemsep{0em}
\item To our best knowledge, PANTHER is the first in jointly modeling both genetic pathways and variants, accounting for their interactions and using groups of interacting pathways as features. 
\item Experiments on both Next Generation Sequencing (NGS) and whole-genome genotyping datasets across diverse diseases demonstrate wide applicability of PANTHER. These experiments show that PANTHER significantly improve accuracies over multiple state-of-the-art comparison models. 
\item Our feature analysis shows that PANTHER can identify insights on disease genetic risks from interacting molecular mechanisms. PANTHER has a GPU implementation. 
\end{itemize}

\section{Related Work}
\textbf{Nonnegative Matrix Factorization (NMF).} NMF is a highly effective unsupervised feature learning tool for single modal data structure exploration (see review~\cite{wang2012nonnegative}). The field of genetic medicine sees many applications of NMF, e.g., clustering cancer patients' somatic mutations and differentiating associations with different cancer types~\cite{alexandrov2013signatures,zeng2019cancer}. In this paper, we use NMF on genetic variants and pathways, both separately and in conjunction, as baselines. 

\textbf{Group Guided and Hierarchical Linear Models.} There are emerging statistical and machine learning methods that have been applied to incorporate gene pathway information into feature selection e.g., ~\cite{sokolov2016pathway,zhou2013polygenic}, including group guided Lasso models e.g., grpreg~\cite{breheny2015group}, and Bayesian hierarchical generalized linear models e.g., brms~\cite{burkner2018advanced}. In this paper, we use state-of-the-art implementations of grpreg and brms on genetic variants or pathways or both as baselines.

\textbf{Nonnegative Tensor Factorization (NTF).} Generalizing the matrix factorization, tensor factorization (see review~\cite{kolda2009tensor}) has recently gained traction in biomedical applications, e.g., genetic association study, population genetics, and transcriptomic analysis (see review~\cite{luo2016tensor}). In the broad field of genetics, most NTF applications are modeling interactions between atomic features only, e.g., between individual gene loci~\cite{hore2016tensor}, between multiple types of bacteria biomarkers~\cite{ozcaglar2011sublineage}. There lacks tensor models for learning groups of higher-order genetic pathways as features to maximize NTF's utility. 

\textbf{Constrained NTF} Incorporating constraints or guidance in NTF has seen increasing development over the past few years. In computational phenotyping, Rubik~\cite{wang2015rubik} adapted CANDECOMP-PARAFAC (CP) factorization to enforce sparsity constraints, accounted for a bias tensor in addition to phenotype tensors, and incorporated medical knowledge. Later, SUSTain~\cite{perros2018sustain} extended real-valued matrix and tensor factorizations to data where values are integers. Moreover, SURF~\cite{he2018boosted} proposed a sparse and low-rank supervised tensor regression model to relate outcomes to a feature tensor. LOgical factorisation Machines (LOM)~\cite{rukat2018probabilistic} proposed a probabilistic Boolean tensor decomposition method using scalable sampling-based posterior inference. This year, LogPar~\cite{yin2020logpar} modeled a tensor with missing data as a binary tensor with Bernoulli distribution parameterized by an underlying real-valued tensor and added uniqueness and smoothness constraints to aid interpretability. In addition, TASTE~\cite{afshar2020taste} combines the
constrained PARAFAC2 model with NMF to jointly model both varying clinical information and demographic information. 

Solving for NTF with growing number of constraints usually requires increasingly complex calculation steps. Most of the state-of-the-art NTF methods rely on Alternating Least Squares (ALS)~\cite{kolda2009tensor} or Alternating Direction Method of Multipliers (ADMM)~\cite{boyd2011distributed}. They are often not designed to run on GPU to benefit from the speedup with GPU parallelization. Moreover, changing the constraints will often result in deriving new optimization procedures, as exemplified by the evolution of constrained NTF models reviewed above.

In genetic medicine, however, there lacks exploration of constrained NTF methods to suit the need for genetic disease risk identification. There are major unmet needs regarding existing factorization methods on the problem of jointly modeling genetic pathways and variants to group interacting pathways and identify co-functioning molecular mechanisms of diseases.

\section{Methods}
We develop an unsupervised feature learning framework that can lead to both more accurate and more interpretable machine learning models for genetic medicine. The model uses genetic pathways to augment NTF and learn interacting groups of pathways. 

\subsection{PANTHER workflow}
\begin{figure*}[t]
    \centering
    \includegraphics[width=1\textwidth]{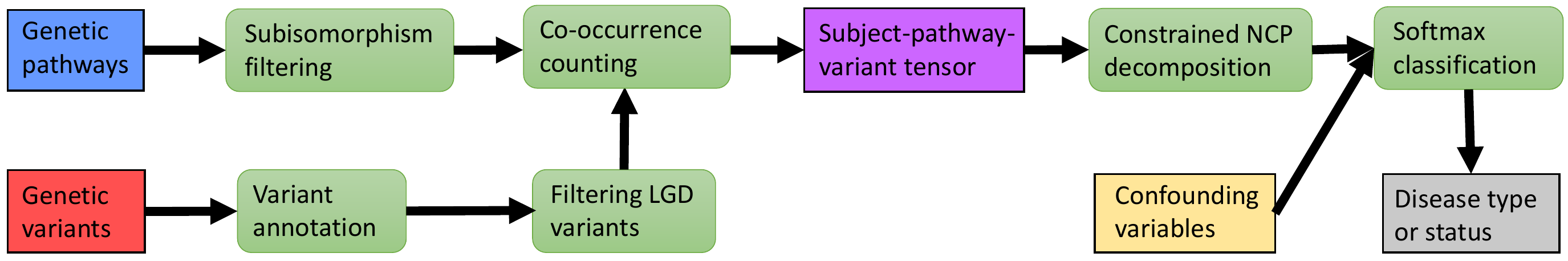}
    \caption{PANTHER workflow. Data are denoted as square boxes, and steps are denoted as round corner boxes. We first annotate the genetic variants and then keep those variants that are deleterious. We filter small genetic pathways that are part of larger pathways (subisomorphism filtering). We then devise a co-occurrence counting scheme to construct the tensor. We further perform constrained Nonnegative CP factorization. In the final disease classification step, the learned subject factor matrix and the confounding matrix are concatenated as features.}
    \label{fig:workflow}
\end{figure*}

Fig.~\ref{fig:workflow} outlines PANTHER's workflow. Table~\ref{tab:notation} defines core symbols that are used in the rest of the paper. PANTHER jointly models higher-order features (genetic pathways) and atomic features (genetic variants). For atomic features, we annotate the genetic variants and retain the deleterious ones. We filter small genetic pathways that are part of larger pathways. A novel heuristic for co-occurence counting is then proposed to generate the \textit{subject} $\times$ \textit{pathway} $\times$ \textit{variant} tensor. Constrained Nonnegative CP factorization is then performed on the tensor to learn the subject factor matrix, which is combined with confounding variables to classify disease types. Detailed steps are explained next.

\begin{table} 
\centering
\begin{tabular}{lrr}  
\toprule
Notation  & Definition \\
\midrule
$\ten{X}$       & \textit{Subject} $\times$ \textit{pathway} $\times$ \textit{variant} tensor      \\
$\mat{S}$    & Subject factor matrix      \\
$\mat{P}$   & Pathway factor matrix      \\
$\mat{V}$   & Variant factor matrix      \\
$\mat{S}_r$ & $r^{th}$ column of $\mat{S}$ \\
$S$ & Subject number \\
$P$ & Pathway number \\
$V$ & Variant number \\
$R$ & Latent group number \\
$\outprod$ & Outer product \\
\bottomrule
\end{tabular}
\caption{Core symbols used in the paper.}
\label{tab:notation}
\end{table}

\subsection{Genetic variant annotation and selection}
The genetic variant input consists of Variant Call Format (VCF) files. A VCF row describes characteristics of a Single Nucleotide Polymorphism (SNP), insertion/deletion (indel) or other genetic variants, their chromosomes and base-pair locations, and their occurrences (e.g., in 0, 1 or both strands). ANNOVAR~\cite{wang2010annovar} is used to annotate each genetic variant and provide comprehensive information, including the hosting gene, the function, the predicted pathogenicity, the minor allele frequency, and the phenotype associations of the genetic variant. We follow conventions to treat low quality variants as 0 variants except for the case of LogPar as baseline where the model treats them as missing data.

Reference mis-annotation may result in wrongly called variants, we use the  exome dataset from Exome Aggregation Consortium (ExAC) to filter out such mis-annotations using the variant frequency aggregated from many large-scale sequencing projects~\cite{lek2016analysis}. Recent debates have been over the issue of whether rare and/or common genetic variants should be the primary target for disease association and the consensus is growing towards including both~\cite{gibson2012rare}. Thus, we retain rare and modestly common variants whose allele frequencies are less than or equal to 90\% as observed from the ExAC cohort. We select deleterious variants for downstream analysis, which include splice site alterations, nonsense variants, and frame-shift insertions/deletions. Many genetic variants produce the same amino acids due to codon redundancy, selecting deleterious variants that are likely harmful leads to more focused study and is an important step in genetic analysis. 

\subsection{Collecting and pruning genetic pathways}
The REACTOME database~\cite{croft2010reactome} is used to obtain a comprehensive collection of human genetic pathways. The biological pathways in REACTOME are expert curated from scientific literature. Larger pathways can sometimes contain smaller pathways, which leads to information redundancy. In addition, spurious associations with disease phenotypes can occur for small pathways due to large single-gene or single-SNP effects~\cite{holmans2010statistical}. To address this issue, we filter the smaller pathway when it is part of, or formally, subisomorphic to, a larger pathway. Let $G_s=(V_s,E_s,l_s)$ and $G=(V,E,l)$ be two pathways (encoded as graphs), where $V$ ($V_s$) is the set of nodes, $E$ ($E_s$) is the set of edges and $l$ ($l_s$) is the function that labels nodes and edges. For $G_s$ to be subisomorphic to $G$, we must have: 
\begin{itemize}
    \item There is a mapping $f: V_s \rightarrow V $ such that   
    \begin{align}
    l_s (v) = l(f(v)) \; \text{for} \; v \in V_s, f(v) \in V
    \end{align}
    \item $\forall (v_1,v_2) \in E_s, \exists (f(v_1),f(v_2)) \in E$ such that 
    \begin{align}
    l_s (v_1,v_2 )=l(f(v_1 ),f(v_2 ))
    \end{align}
\end{itemize}
For this work, we simplify the definition of subisomorphism to containment in that $f$ now becomes the identity mapping and $l$ and $l_s$ need to produce the same labels for the nodes and edges shared by the graphs. In addition, multiple heuristics are used to prune the comparisons of subisomorphism that are unnecessary, as in Algorithm~\ref{alg:alg-subiso} (e.g. size heuristic in line 2 and subset precheck in line 5). Note that PANTHER does keep many small pathways, as long as they are not contained in a larger pathway.

\begin{algorithm}[tb] 
\caption{Subisomorphism detection for pathways}
\label{alg:alg-subiso}
\textbf{Input}: $\mathcal{S}$ – set of pathways ~~ \\
\textbf{Output}: $H$ – hash table of discovered subisomorphisms

\begin{algorithmic}[1] 
\STATE Let $H=\{\}$
\STATE Stable sort $\mathcal{S}$ in ascending order of number of nodes
\FOR{ $i = 1$ to length($\mathcal{S}$) $- 1$}
\FOR{ $j = i+1$ to length($\mathcal{S}$)}
\IF {nodes($\mathcal{S}[i]$) $\subset$ nodes($\mathcal{S}[j]$)}
\IF {subisomorphism($\mathcal{S}[i], \mathcal{S}[j]$)}
\STATE $H[\mathcal{S}[i]]=\mathcal{S}[j]$ 
\ENDIF
\ENDIF
\ENDFOR
\ENDFOR
\end{algorithmic}
\end{algorithm}

\subsection{Subject $\times$ Pathway $\times$ Variant Tensor}
\begin{figure*}[t]
    \centering
    \includegraphics[width=0.8\textwidth]{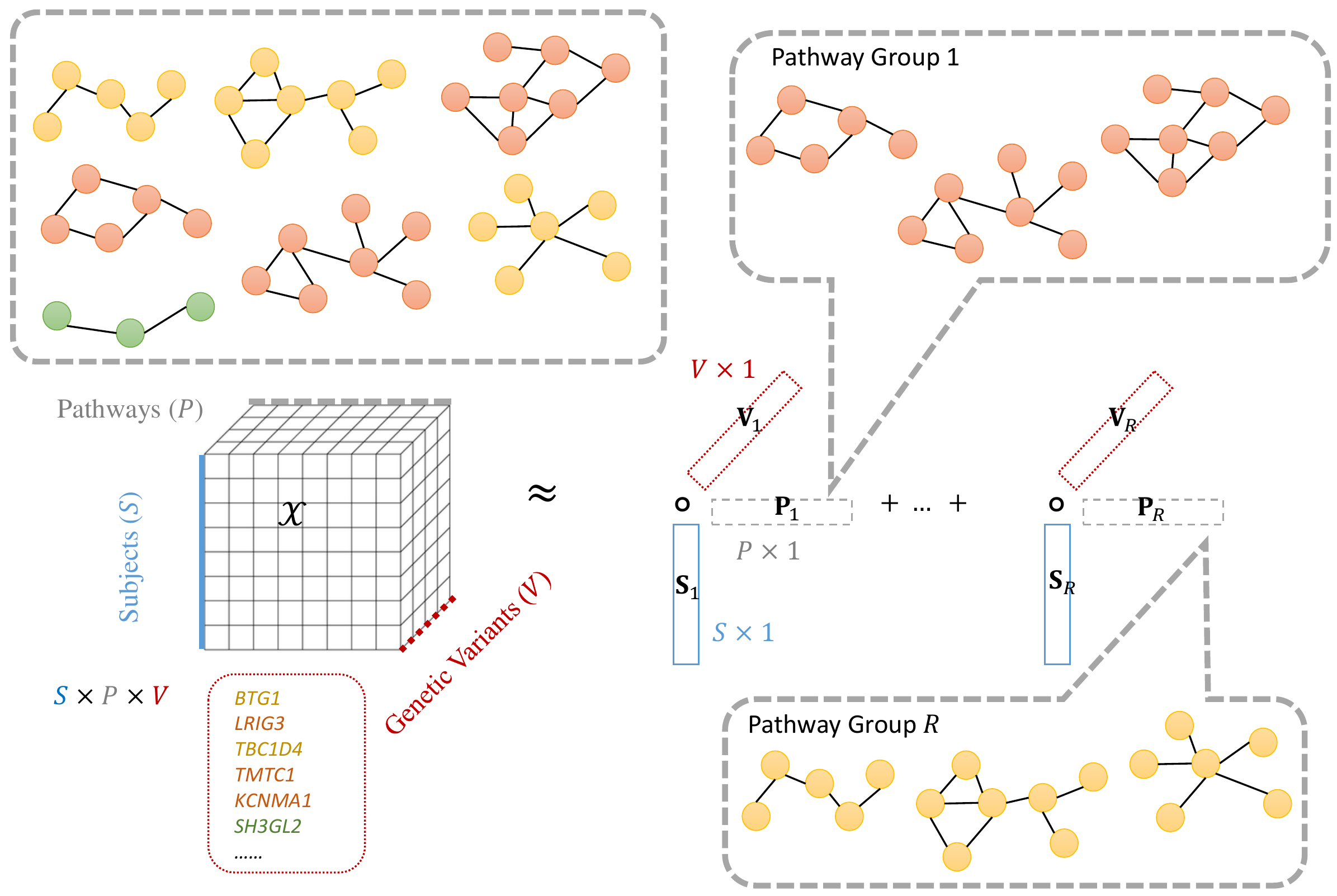}
    \caption{PANTHER's tensor factorization illustration for jointly modeling genetic pathways and variants.}
    \label{fig:panther}
\end{figure*}

Fig.~\ref{fig:panther} describes the construction of the \textit{subject} $\times$ \textit{pathway} $\times$ \textit{variant} tensor $\ten{X}$. The tensor entries record the co-occurrence counts. Using the high dimensional variants directly as one mode of the tensor may result in impractically large memory consumption. Thus, we aggregate the count of genetic variants to the genes they affect and abuse terminology by referring “gene” and "variant" exchangeably in the following text. In Fig.~\ref{fig:panther}, the factor matrix \mat{S} is the \textit{subject} $\times$ \textit{subject group} matrix, \mat{P} the \textit{pathway} $\times$ \textit{pathway group} matrix, \mat{V} the \textit{gene} $\times$ \textit{gene group} matrix.  Let $\mat{M}$ be the \textit{subject} $\times$ \textit{gene} matrix, where $\mat{M}_{i,k}$ denotes how many variants occur in gene $k$ in subject $i$. Let $\mat{N}$ be the \textit{subject} $\times$ \textit{pathway} matrix, where $\mat{N}_{i,j}$ denotes how many variants occur in pathway $j$ in subject $i$. The tensor $\ten{X}$ is constructed as in equation~\ref{eq:tensor-str}.

\begin{align}
\ten{X}_{ijk} = 
\begin{cases}
\quad \mat{M}_{i, k} & \text{subject } i \text{ has mutating gene } k \\ 
\quad & \text{that belongs to pathway } j \\
\quad \mat{N}_{i, j} & \text{gene } k \text{ and pathway } j \text{ co-occurs}\\
\quad & \text{in subject } i \text{ and } \mat{N}_{i, j} \le \mat{M}_{i, k}\\
\quad \mat{M}_{i, k} & \text{gene } k \text{ and pathway } j \text{ co-occurs}\\
\quad & \text{in subject } i \text{ and } \mat{N}_{i, j} > \mat{M}_{i, k}\\
\quad 0&  \text{otherwise}
\end{cases}
\label{eq:tensor-str}
\end{align}

Besides where the gene belongs to the pathway, we also account for the case where the subject has some variants that hit a gene and other variants that hit a pathway not containing the gene, thus making them "co-occur". The co-occurence count is defined as the lesser of the counts for the two groups of variants; this avoids simply replicating slices of tensors. There are certainly alternative ways to define co-occurence count, but we observe that the definition in equation~\ref{eq:tensor-str} works well in our experiments. Intuitively, the heuristic counts additional gene-pathway co-occurrence where the patient has some variants that hit a gene and other variants that hit a pathway not containing the gene. This generalized by-patient co-occurrence is motivated by the fact that we currently do not know all possible interactions between genes and pathways, as knowledge is still evolving and knowledge-based co-occurrence can be incomplete. Adding the data-driven co-occurrence helps supplement the incomplete and currently evolving knowledge.

\subsection{PANTHER's factorization step}
As defined in Table~\ref{tab:notation} and used in Fig.~\ref{fig:panther}, $R$ is the latent group number. The following factor matrices combine the corresponding vectors in each mode: 
\begin{gather}
\mat{S} = [\mat{S}_1 \; | \; ... \;  | \;  \mat{S}_r \;  | \;  ... \;  | \;  \mat{S}_R] \in \mathbb{R}^{S \times R} \nonumber\\
\mat{P} = [\mat{P}_1 \; | \; ... \;  | \;  \mat{P}_r \;  | \;  ... \;  | \;  \mat{P}_R] \in \mathbb{R}^{P \times R} \nonumber\\
\mat{V} = [\mat{V}_1 \; | \; ... \;  | \;  \mat{V}_r \;  | \;  ... \;  | \;  \mat{V}_R] \in \mathbb{R}^{V \times R}
\end{gather}
Let $1 \leq r \leq R$, for the $r^{th}$ vectors from matrices $\mat{S},\mat{P},\mat{V}$, their outer product is defined as 
\begin{equation}
\ten{T} = \mat{S}_r \outprod \mat{P}_r \outprod \mat{V}_r
\end{equation}
where the entries of the rank-one tensor are $\ten{T}_{i,j,k} = \mat{S}_{i,r} \mat{P}_{j,r} \mat{V}_{k,r}$. The CP factorization decomposes the tensor $\ten{X}$ into an array of rank-one tensors, which is formulated as
\begin{equation}
\ten{X} \approx \sum_{r = 1}^{R}\mat{S}_r \outprod \mat{P}_r \outprod \vec{V}_r =   \llbracket \mat{S},\mat{P},\mat{V} \rrbracket
\end{equation}
We require the factorization result entries to be nonnegative so that the model can be easily interpreted additively. Informed by the application in genetic medicine, the following constraints are furthered added to the model 
\begin{gather} 
\min_{\mat{S}, \mat{P}, \mat{V}} \norm{ \ten{X} - \llbracket \mat{S},\mat{P},\mat{V} \rrbracket }_F^2 + \lambda_1  \phi(\mat{S}) \nonumber\\ 
\text{s.t. } \mat{S} \ge 0, \mat{P} \ge 0, \mat{V} \ge 0 \nonumber\\
\mat{P} \in \{0\} \cup [\gamma_P, +\infty)^{P \times R} \nonumber\\
\mat{V} \in \{0\} \cup [\gamma_V, +\infty)^{V \times R}
\label{eq:obj}
\end{gather}
where,
\begin{equation}
\phi(\mat{S}) = \norm{\mat{I} - R \cdot \mat{S}^T \mat{S} / \sum \mat{S}^T \mat{S} }_F^2
\label{eq:ortho}
\end{equation}
The downstream classification step will use the subject factor matrix $\mat{S}$ in the feature matrix. In genetic medicine, evaluating the features' effect sizes is a routine requirement, but correlations among features often makes such evaluation inaccurate or biased. Thus, we add the constraint $\phi(\mat{S})$ to impose the orthogonality among features, which is formulated in equation~\ref{eq:ortho}. One typically only needs the feature vectors to be orthogonal without requiring them to have unit norm, so equation~\ref{eq:ortho} is formulated in a scale-free fashion. For clinical applications, sparsity on the learned high-order features will allow focusing on only a few key patterns at a time hence easy interpretation. In this work, we use the parameters $\gamma_P, \gamma_V$ to control the sparsity levels on the pathway and variant factor matrices.

We implement PANTHER's factorization on GPU using Tensorly ~\cite{kossaifi2019tensorly} and PyTorch, and solve its optimization using ADAM~\cite{kinga2015method}. We train the factorization for up to 6000 iterations with early stopping when the validation loss does not decrease from the average of 10 previous consecutive epochs. The motivations for adopting ADAM instead of hand-solving equation~\ref{eq:obj} are two folds: 1) as an autograd optimizer, ADAM allows flexible configurations of constraints (e.g., orthogonality and sparsity on customarily selected matrices) without deriving the gradients from scratch every time; 2) as an SGD optimizer, ADAM can handle large scale problems (e.g., single cell genomics / transcriptomics data) with minibatch. Such choice is practically important and enables rapid iterations of flexibly customized problem formulations and mass experiments when working with genetic medicine data.

\subsection{PANTHER's classification step}
In order to avoid biased model parameter estimation in genetic medicine, one needs to explicitly account for confounding variables. In the PANTHER framework, we use the matrix $\mat{F}$ to capture the variables that are confounding. $\mat{F}$ usually consists of age, gender, and race for genetic medicine applications, and is usually low dimensional. The learned feature matrix $\mat{S}$ and the confounding matrix $\mat{F}$ are then concatenated as input to the following softmax classifier
\begin{equation}
\mat{Z} = \text{softmax}(\mat{W}^{(1)} \; [\mat{S} \: | \: \mat{F}] \: + \: \mat{W}^{(0)})
\end{equation}
We calculate the cross-entropy loss in all classes over all training subjects
\begin{equation}
\mathcal{L} = -\sum_{i \in \mathcal{Y}_{tr}} \sum_{c=1}^C \mat{Y}_{ic} \ln \mat{Z}_{ic}
\label{eq:floss}
\end{equation}
where $\mathcal{Y}_{tr}$ consists of the subjects in the training set, and $C$ is the number of different classes, i.e., unique labels. $\mat{Y}$ is the binary indicator matrix for class labels. The matrices $\mat{W}^{(0)}$ and $\mat{W}^{(1)}$ are the weight matrices and have the number of rows being equal to the number of classes.

\section{Experiment I: Using NGS Data to Predict Cancer Type}
In this experiment, four prevalent cancers were retrieved from The Cancer Genome Atlas (TCGA), including breast cancer, lung cancer, colorectal cancer and prostate cancer. Germline variants are variants inherited from parents and can inform early screening for different cancer types to enable early interventions~\cite{bertelsen2019high}. Thus, we focus on germline variants in this work. Table~\ref{tab:tcgadata} shows the partitioning of the subjects (2545 total), stratified by cancer types, into a training set (1527 subjects total), a validation set (509 subjects total) and a held-out test set (509 subjects total). The filtered pathway and gene numbers are 626 and 684 respectively. 

\begin{table}
\centering
\begin{tabular}{lrrrr}
\toprule
Disease  & Total & Train & Validation  & Test \\
\midrule
Breast cancer & 959  & 575 & 192 & 192  \\
Colorectal cancer & 728  & 437 & 146 & 145   \\
Lung cancer   & 440  & 264 &  88  & 88 \\
Prostate cancer  & 418  & 251 & 83 & 84     \\
\bottomrule
\end{tabular}
\caption{Distribution of the TCGA dataset. Our dataset consists of breast cancer, colorectal cancer, lung cancer and prostate cancer that are the four most prevalent cancer types. We use a 6:2:2 ratio to split the dataset into a training set, a validation set and a test set.}
\label{tab:tcgadata}
\end{table}

\textbf{Parameters.} When using nonnegative matrix and tensor factorizations to identify latent groups of features used by the model, we empirically determine the number of groups $R$. For all NMF and tensor methods except for SURF, we use the validation set to tune R from choices between 50 and 500 (at increments of 50). SURF by default scans $R$ from 1 to 500 using a step size 1. One typically does not need to set such a small step size. From our correspondence with SURF’s senior author, its step size 1 is required for efficient computation using their specific deflation method that traces the solution path. We also experimented with step size 50 for SURF, and the best accuracies have very small changes (<0.005) with $R$ moving to adjacent choice on the new search grid. For PANTHER, $\lambda_1$ is tuned using the validation set from choices including (0.01, 0.1, 1, 10, 100) and is set to 1. The sparsity thresholds $\gamma_P, \gamma_V$ are chosen by following the sensitivity analysis described in~\citep{wang2015rubik} and are set to 0.01 and 0.001 respectively. 

Comparison tensor methods also have multi-part loss functions that are weighted by $\lambda’$s. They are tuned on validation set using choice grids according to respective papers, or from the default (0.01, 0.1, 1, 10, 100) if unspecified. For classification with pLR, NMF, and tensors, we tune the parameter that balances the empirical error and model complexity using the validation set from choices of (0.01, 0.1, 1, 10, 100). 

\textbf{Settings.} Regarding comparison model configurations, for Bayesian hierarchical model brms, we treat confounding factors as upper level features for $\text{brms}_{\text{gene}}$ and $\text{brms}_{\text{pathway}}$, and add pathways to upper level features for $\text{brms}_{\text{gene+pathway}}$. The model grpreg requires user-specified non-overlapping groups. We partition pathways into groups that do not share genes, and in turn partition genes into groups based on their memberships in pathway groups. The corresponding $R$ values in Table~\ref{tab:tcga-acc} are in fact those group numbers. For comparison NTF models, in LogPar we treat low quality variants as missing data instead of 0 variants. TASTE lets us use subject demographics to help with factorization. SURF is applied as a supervised NTF model. For all models except for deterministic penalized logistic regression, we run 10 random initializations and use the validation set to select the best initialization.

\begin{table}
\centering
\begin{tabular}{lrr}  
\toprule
Model  & $R$ & Test Accuracy  \\
\midrule
$\text{pLR}_{\text{gene}}$  & - & 0.8016      \\
$\text{pLR}_{\text{pathway}}$ & - & 0.7701 \\
$\text{pLR}_{\text{gene+pathway}}$ & - & 0.7682 \\
\midrule
$\text{brms}_{\text{gene}}$   & - & 0.8016      \\
$\text{brms}_{\text{pathway}}$   & - &   0.7642   \\
$\text{brms}_{\text{gene+pathway}}$   & - &  0.8173     \\
\midrule
$\text{NMF}_{\text{gene}}$         & 400 & 0.8173      \\
$\text{NMF}_{\text{pathway}}$  & 400 & 0.7819 \\
$\text{NMF}_{\text{gene+pathway}}$ & 100 & 0.7957 \\
\midrule
$\text{grpreg}_{\text{gene}}$   & 66 & 0.6306      \\
$\text{grpreg}_{\text{pathway}}$   & 66 & 0.8153      \\
$\text{grpreg}_{\text{gene+pathway}}$   & 131 &   0.8016    \\
\midrule
Rubik    & 100 & 0.8035      \\
SUSTain   & 450 & 0.7839      \\
SURF & 50 & 0.7466 \\
LogPar & 350 & 0.6031  \\
TASTE & 200 & 0.8369  \\
LOM & 100 &  0.7505  \\
PANTHER & 200 & \bf{0.8644} \\
\bottomrule
\end{tabular}
\caption{Cancer type prediction accuracy on the TCGA test dataset. PANTHER significantly outperforms all of the state-of-the-art comparison models ($p < 0.05$, permutation test). pLR: penalized logistic regression.} 
\label{tab:tcga-acc}
\end{table}

\textbf{Test Performance.} Table~\ref{tab:tcga-acc} presents the test accuracy of each model. PANTHER performs the best and is significantly better than all baseline models ($p<0.05$ by random permutation test). For more detailed performance analysis, we note that  NMF on gene count matrix is modestly better than penalized logistic regression models on the gene and/or pathway count matrices. For NMF, direct concatenation of the gene count matrix with the pathway count matrix does not lead to positive change in accuracy. The group guided model grpreg and the Bayesian hierarchical model brms do not give better results than NMF, indicating that simply adding pathways into another feature hierarchy does not improve accuracy. For NTF baselines, Rubik and SUSTain do not further improve accuracy from NMF models, SURF, LOM and LogPar actually have considerable decrease in accuracy, TASTE outperforms NMF by a margin. But all are significantly outperformed by PANTHER by a notable margin. 

Pathway features are less sparse than gene features and capture more correlation between patients. On the other hand, a gene feature is more likely to show difference across disease types than a pathway, which sometimes can be an artifact due to sparsity. Flat models like logistic regressions may view pathways and genes as competing features and lack effective ways to harmonize them. When we use dimensionality reduction method, group guided method and hierarchical method, pathways start to contribute more positively together with genes. Certain tensor models, with their multi-modal formulations, are observed to strike better balances between capturing correlation, maintaining discriminating power and reducing sparsity, hence improving performance. PANTHER’s co-occurrence counting heuristic further improves the formulation, and gives best results when combined with properly motivated constraints in equation~\ref{eq:obj}.

PANTHER also runs much faster than state-of-the-art NTF models. For example, with $R=200$, PANTHER, Rubik, TASTE run in 13min, 59min, >12 hours respectively (GPU V100, CPU 24 cores). These results indicate that various constraints and supervision on NTF may not fit the task in a different domain (e.g. from phenotyping to genetics) and it is important to use the precise set of properly motivated constraints. The trial and err entailed in identifying these constraints again underscores the benefit of our choice of the generic optimizer ADAM (vs. hand derived gradients) that allows for rapid iterations. 

\begin{figure}[t]
\centering
\subfigure[$\text{pLR}_{\text{gene}}$]{
\label{fig:tcgavis:a} %
\includegraphics[height = 27 mm]{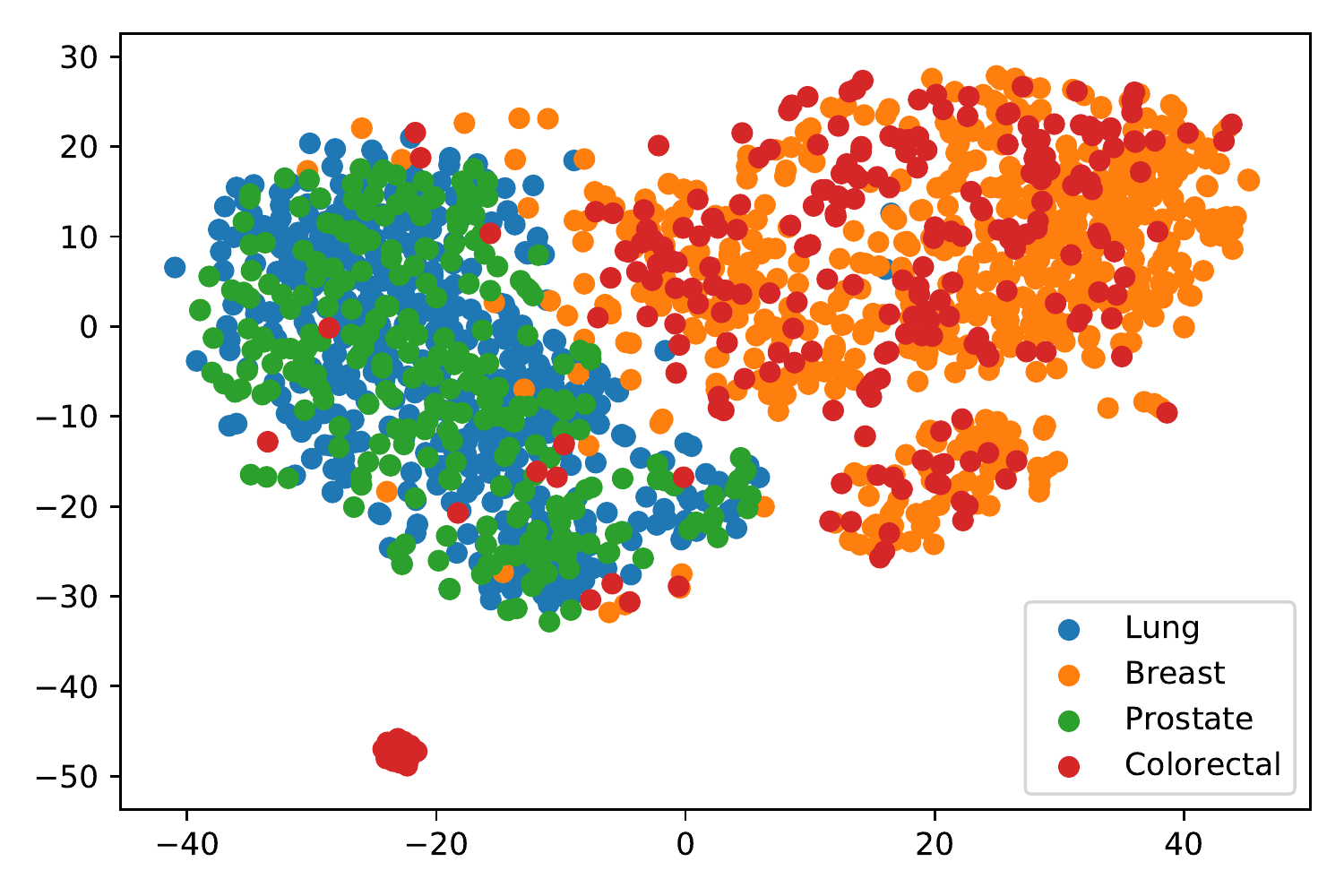}} 
\subfigure[Rubik]{
\label{fig:tcgavis:b} %
\includegraphics[height = 27 mm]{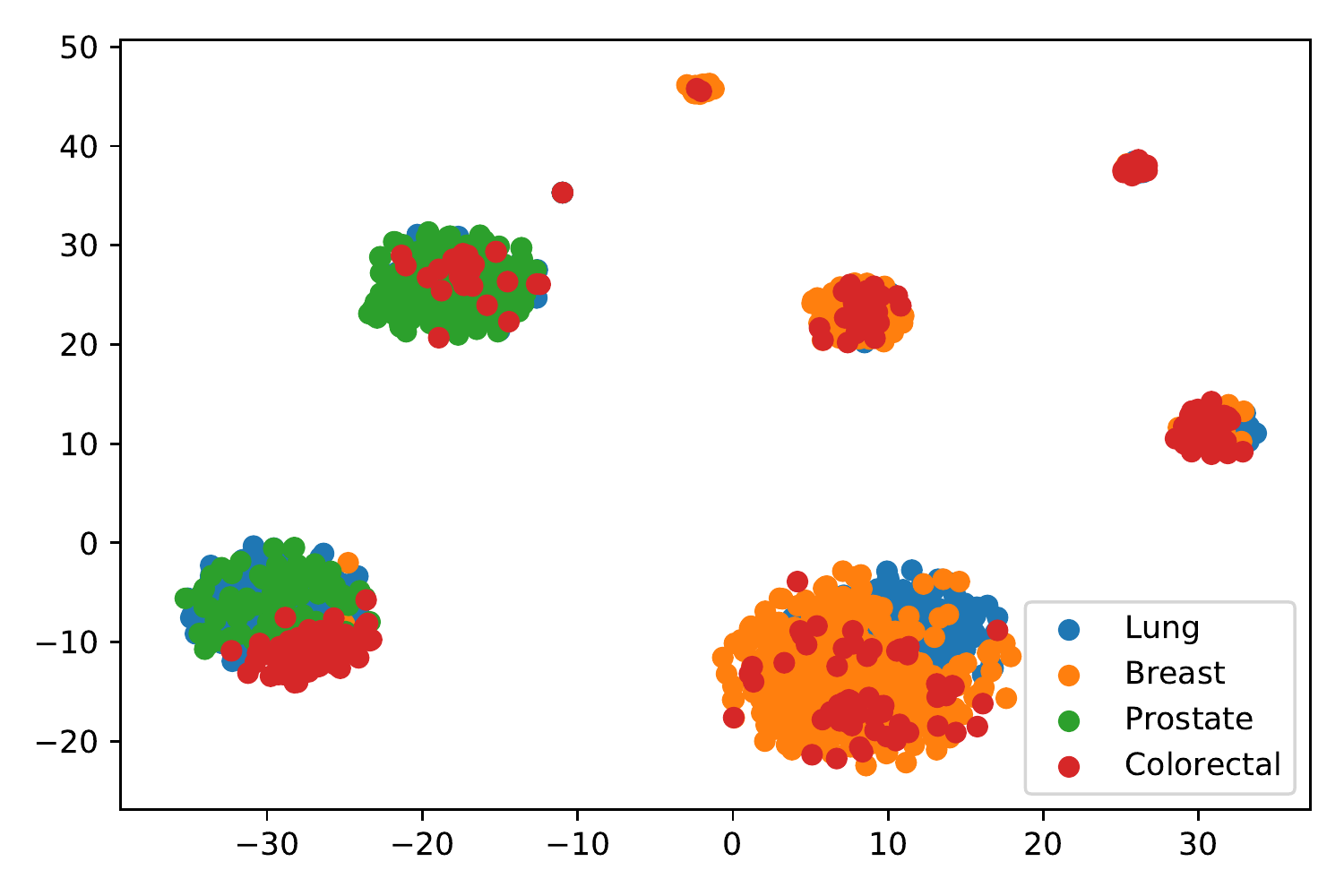}}
\subfigure[TASTE]{ 
\label{fig:tcgavis:c} %
\includegraphics[height = 27 mm]{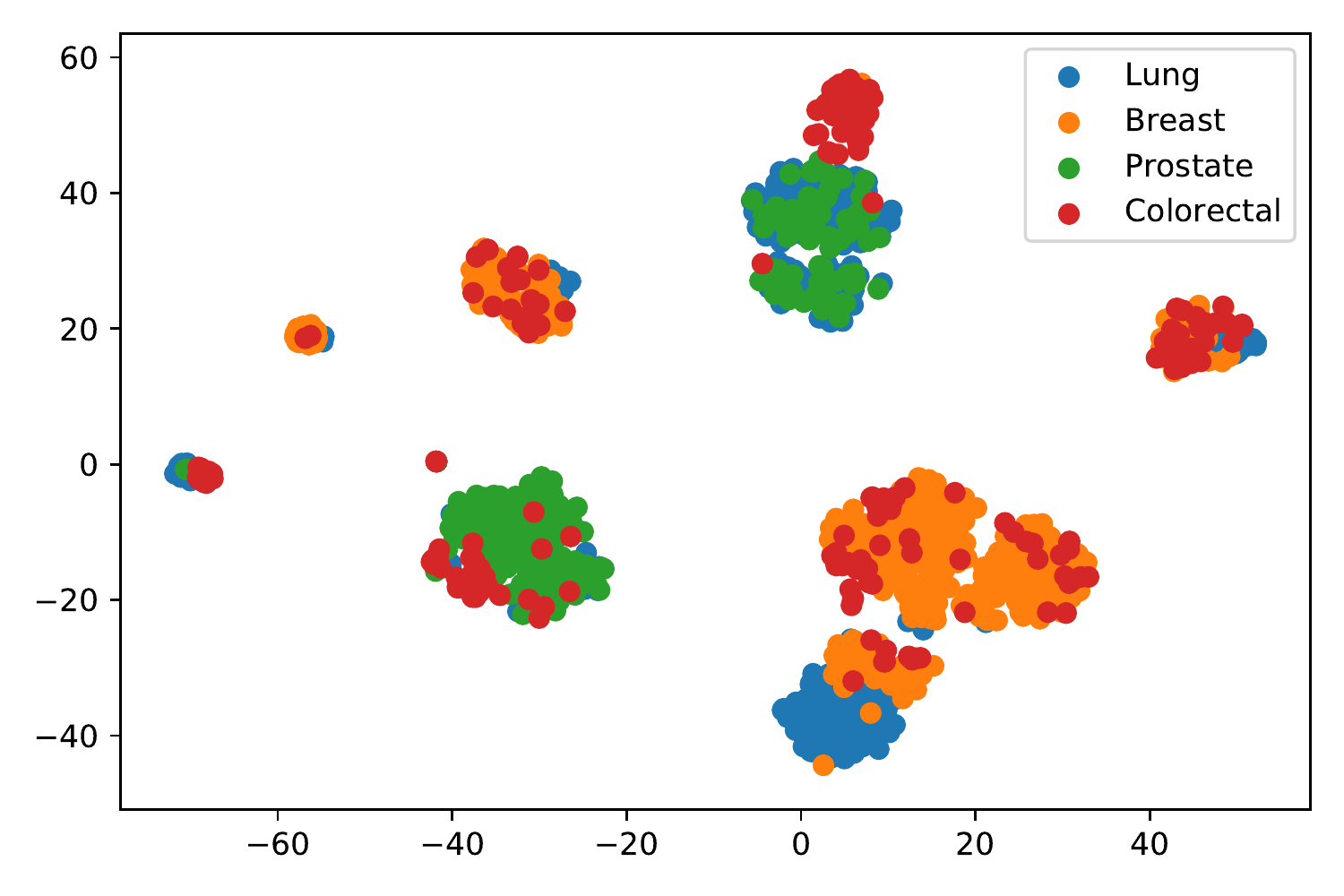}} 
\subfigure[PANTHER]{
\label{fig:tcgavis:d} %
\includegraphics[height = 27 mm]{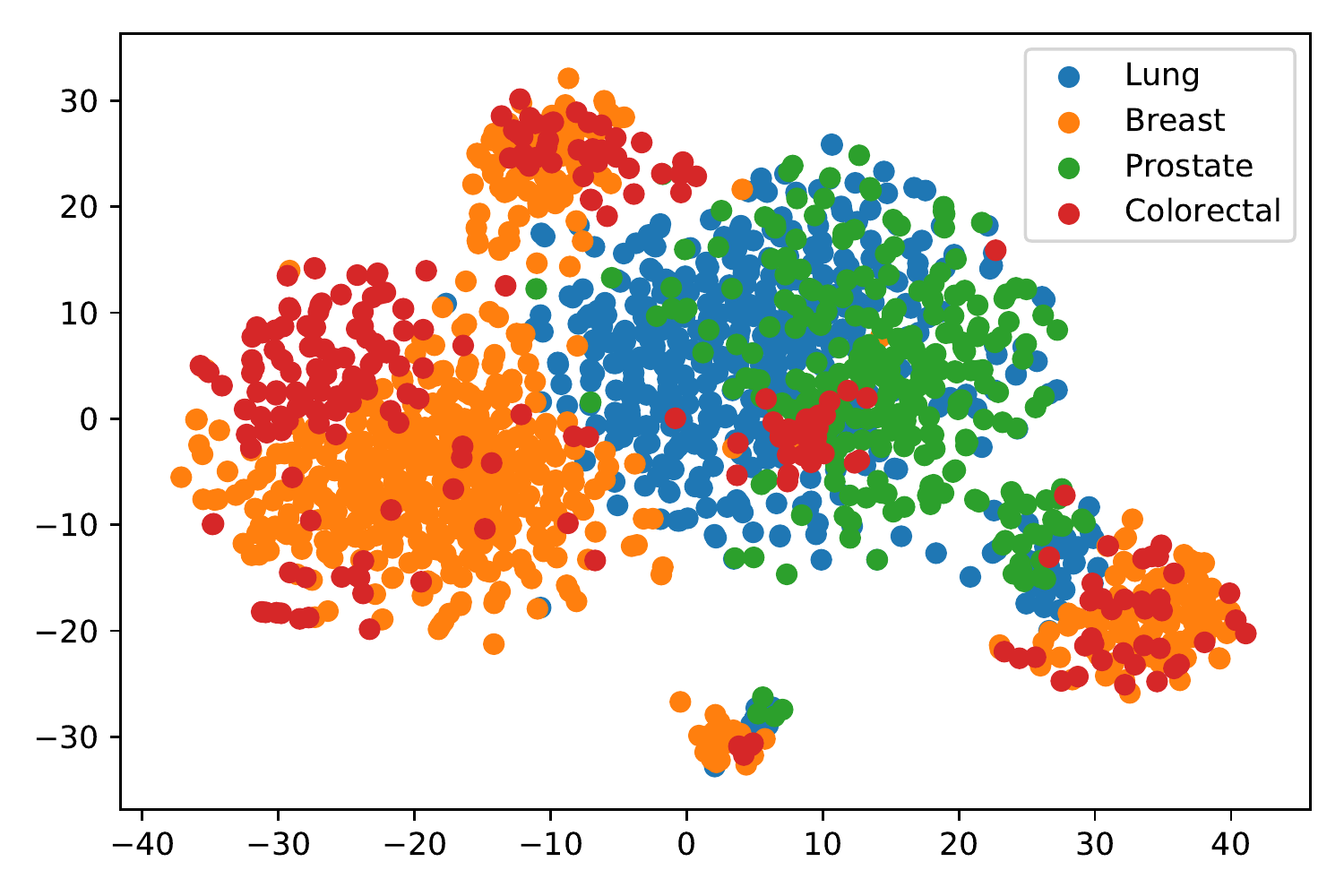}}
\caption{The t-SNE visualization of the learned subject features in TCGA training dataset.}
\label{fig:tcgavis}
\end{figure}

\textbf{Visualization.} Visualizing the subjects' features learned by PANTHER and comparison models allows for a direct illustration of their respective effectiveness. We use t-SNE~\cite{maaten2008visualizing} and show TCGA visualization. Fig.~\ref{fig:tcgavis} displays t-SNE plots of the representation of subjects learned by representative models including penalized logistic regression, Rubik, TASTE and PANTHER from Table~\ref{tab:tcga-acc}. Fig.~\ref{fig:tcgavis} suggests that PANTHER produces more discriminative subject representations, compared with penalized logistic regression on gene count matrix and state-of-the-art constrained tensor factorization models.

\begin{table*}[t]
\footnotesize
\centering
\renewcommand{\arraystretch}{1.2}
\label{table_1}
\caption{Identified genetic pathway groups for the four prevalent types of cancer.}
\begin{tabular}{c|c|c|c}
\hline

Breast & Colorectal & Lung & Prostate \\
\hline
Post-chaperonin tubulin folding  & SMO activation & HIF-$\alpha$ hydroxylation  & Oxidative Stress Induced Senescence\\
PIP synthesis & Wnt signaling & FGFR2 signaling & FZD regulation \\
TP53 regulation & AURKA Activation & piRNA biogenesis  &  PKMTs methylate histone lysines \\
FGFR2 signaling &  $\beta$-catenin formation  & E2F transcription & TRP channels\\
piRNA biogenesis & $\rho$ GTPase cycle &  ERCC6 and EHMT2 regulation  & PLC-$\beta$ mediated events\\
\hline
\end{tabular}
\label{tab:tcga-pathway}
\end{table*}

\textbf{Discovering and interpreting pathway groups.} PANTHER allows the identification of the top features for different cancers in the form of pathway groups. Regarding a cancer class $c$, its corresponding softmax weights $\mat{W}_c^{(1)}$ are ranked in descending order  and the pathway group index corresponding to the highest weight, say $r$, is selected. The coefficients in the pathway group vector $\mat{P}_r$ are in turn ranked to identify the indices of top pathways linked to the cancer class $c$. Table~\ref{tab:tcga-pathway} shows the identified pathway groups for the four types of cancer we study. It is worth noting that most of the identified pathways are key events that are innate in the oncogenesis of one or more of the four prevalent cancer types, e.g., fibroblast growth factor receptor (FGFR2) signaling pathway~\citep{campbell2016fgfr2}, ubiquitination regulation of human frizzled (FZD) genetic pathway~\citep{ueno2013frizzled}, as well as piRNA expression in breast and lung cancers~\citep{chalbatani2019biological}.  

Interestingly, PANTHER also discovers interactions between pathways that offer molecular mechanistic insights and novel therapeutic opportunities in individual cancer types. For example, regarding the top breast cancer pathway group, the disruption in phosphatidylinositol phosphate (PIP) synthesis pathway and the inhibition of FGFR2 signaling, together with TP53 aberrations, are potential oncogenic drivers of Triple-negative breast cancers~\citep{liu2018identifying}. Regarding the top colorectal cancer pathway group, disruption to AURKA pathway acts together with Wnt signaling pathway in the onset and progression of colorectal cancer~\cite{jacobsen2018aurora}. Regarding the top lung cancer pathway group, comprehensive gene expression analysis revealed that a sustained disruption of the E2F transcription pathway was accompanied by impaired HIF-1$\alpha$ hydroxylation pathway. This suggests a novel therapeutic opportunity for palbociclib in lung cancer currently treated with taxane based chemotherapy as standard of care~\cite{cao2019combining}. Regarding the top prostate cancer pathway group, innate oxidative stress is associated with prostate cancer development, progression and the response to therapy, and the epigenetic events such as hypermethylation by PKMTs profoundly reduce the cells’ antioxidative capacity and promote prostate cancer aggressiveness~\citep{paschos2013oxidative,he2012targeting}. The above insights indicate that in addition to improving predictive modeling performance, PANTHER can effectively learn grouped genetic pathways as features, and enable the interpretation of the learned model at the level of molecular mechanisms. In particular, PANTHER links together higher-order features by grouping multiple co-functioning pathways, in order to allow collective assessment of their contribution to disease onset and progression. 

\section{Experiment II: Predicting Hypertension Risks Using Whole-Genome Genotyping Data}
Although the name is suggestive of a single disease entity, primary hypertension is a heterogeneous collection of conditions that have varying pathophysiology and come with different etiologies. Primary hypertension has complex genetic risk factors. However, in the field of genetic medicine, existing hypertension studies have typically analyzed individual variants one at a time in order to assess their effect sizes (see~\cite{hypertension2015lancet} for a review). In this experiment, we apply PANTHER to whole-genome genotyping data that is more widely available than NGS, to further demonstrate PANTHER's wide utility.

\begin{table}
\centering
\begin{tabular}{lrrrr}
\toprule
Hypertension  & Total & Train & Validation  & Test \\
\midrule
No & 915  & 549 & 183 & 183  \\
Mild & 622  & 372 & 125 & 125   \\
Severe    & 565  & 339 &  113  & 113 \\
\bottomrule
\end{tabular}
\caption{Statistics of hypertension experiment data. The table includes the distribution of the hypertension subtypes: non-hypertensive, mild hypertensive and severe hypertensive. We use a 6:2:2 ratio to split the dataset into a training set, a validation set and a test set.}
\label{tab:hpgdata}
\end{table}

Our experiment uses a cohort consisting of subjects that were enrolled in a multi-site study that recruited hypertensive individuals and normotensive controls in order to understand hypertension genetic risk factors~\cite{williams2000nhlbi}. Whole-genome genotyping data were collected for both Caucasian and African American participants. 

In this experiment, we use the subjects' clinically measured hypertension subtypes as outcomes including non-hypertensive, mild hypertensive, and severe hypertensive as defined in the JNC VI guideline. We partitioned the subjects (2102 total) according to a 6:2:2 ratio, stratified by hypertension subtypes, into a training set (1260 subjects total), a validation set (421 subjects total) and a held-out test set (421 subjects total), as in Table~\ref{tab:hpgdata}. The filtered pathway and gene numbers are 592 and 717 respectively.

\textbf{Parameters and Settings.} We tune the parameters using the validation set, similarly as in the cancer experiment. $R$ is chosen from choices between 50 and 500 (at increments of 50). For PANTHER, $\lambda_1$ is set to 10 from choices of (0.01, 0.1, 1, 10, 100). The sparsity thresholds $\gamma_P, \gamma_V$ are chosen by following the sensitivity analysis described in~\citep{wang2015rubik} and are set to 0.001 and 0.0001 respectively. The comparison model settings and parameter tuning processes also follow the cancer experiment.

\begin{table}
\centering
\begin{tabular}{lrr}  
\toprule
Model  & $R$ & Test Accuracy  \\
\midrule
$\text{pLR}_{\text{gene}}$  & - & 0.5938     \\
$\text{pLR}_{\text{pathway}}$ & - & 0.5986 \\
$\text{pLR}_{\text{gene+pathway}}$ & - & 0.5819 \\
\midrule
$\text{brms}_{\text{gene}}$   & - & 0.4465      \\
$\text{brms}_{\text{pathway}}$   & - &  0.5416    \\
$\text{brms}_{\text{gene+pathway}}$   & - &  0.4490     \\
\midrule
$\text{NMF}_{\text{gene}}$         & 200 & 0.6390      \\
$\text{NMF}_{\text{pathway}}$ & 350 & 0.5653 \\
$\text{NMF}_{\text{gene+pathway}}$ & 250 & 0.6318 \\
\midrule
$\text{grpreg}_{\text{gene}}$   & 75 & 0.6200       \\
$\text{grpreg}_{\text{pathway}}$   & 75 & 0.5962      \\
$\text{grpreg}_{\text{gene+pathway}}$   & 149 &  0.6010     \\
\midrule
Rubik    & 50 & 0.6437      \\
SUSTain   & 250 & 0.6176      \\
SURF & 150 & 0.3539  \\
LogPar & 150  & 0.6532  \\
TASTE & 50 &  0.6532  \\
LOM & 50 & 0.6508 \\
PANTHER & 450 & \bf{0.6888} \\
\bottomrule
\end{tabular}
\caption{Hypertension type classification accuracy on the held-out test dataset. PANTHER significantly outperforms all comparison models ($p < 0.05$ permutation test). pLR: penalized logistic regression.}  
\label{tab:hpg-acc}
\end{table}

\textbf{Test Performance.} Table~\ref{tab:hpg-acc} shows that PANTHER performs the best and significantly better than all state-of-the-art models ($p<0.05$, permutation test). Detailed comparison suggests similar trends among models as in the cancer experiment with the following differences. NMF on gene count matrix gives more improvements from penalized logistic regression models than group guided and Bayesian hierarchical models. For NTF models, Rubik, LOM, LogPar and TASTE further improve accuracy from NMF models. Still, all are significantly outperformed by PANTHER by a notable margin, reinforcing our intuitions from the cancer result analysis. Due to space limitation, we omit the top pathway group analysis and t-SNE visualization for the hypertension experiment. 

\section{Discussion and Future Work}
In addition to being significantly more accurate and more interpretable, PANTHER uses ADAM optimizer thus works well with minibatch, scales well to large scale problems, and allows rapid iterations of models and mass experiments. On the other hand, our work comes with several limitations. In particular, we only investigated CP factorization scheme. We did preliminary exploration on the constrained Tucker factorization scheme that allows different group numbers in different modes. However, our exploration suggested that learning the constrained Tucker factorization likely requires a much larger dataset that is yet existent in the genetics field. In addition, we did not use the confounding variables to guide the tensor construction and factorization steps. Using confounding variables to guide these steps is technically possible, but may introduce confounding information in the derived features and make downstream analysis more complicated. Care needs to be taken and this will be our future work. We also plan to experiment with end-to-end training of factorization and classification with multiple objectives. Our previous experiments on end-to-end training with matrix factorization achieved promising results, but the extension to tensor factorization will be considerably more complex future work~\cite{pmlr-v126-luo20a}. 

\section{Conclusions}
We proposed the novel PANTHER framework: Pathway Augmented Nonnegative Tensor factorization for HighER-order feature learning for genetic medicine, designed for learning groups of interacting genetic pathways to represent co-functioning molecular mechanisms. Our experiments showed that PANTHER is more accurate and more interpretable than state-of-the-art comparison models when being applied to predict disease types. Genetic pathways directly correspond to molecular mechanisms, which are more informative than individual genes. The ensuing genetics motivated constrained tensor factorization step can identify groups of genetic pathways that reflect co-functioning molecular and disease mechanisms. This not only improves accuracy but also enables interpretation at the level of high-order features that are more natural to clinical reasoning. We compared PANTHER with multiple state-of-the-art constrained matrix and tensor factorization models, as well as group guided and Bayesian hierarchical models. PANTHER significantly ($p < 0.05$, permutation test) outperforms all the comparison models in accuracy. We performed detailed feature analysis of the cancer pathway groups learned by PANTHER, which revealed genetic medicine insights on co-functioning molecular mechanisms being differentially linked to different types of cancer.

\section*{Ethics Statement}
It is important for the AI/ML community to continue being informed about the problems arising in critical application domains such as healthcare and genetic medicine where massive amounts of data are being rapidly generated. More specifically, it is important to understand the need for improved problem formulations, the need for enabling rapid iterations of models and mass experiments without re-deriving the model solutions from scratch due to changed objective functions, the need for better interpretability by grouping interacting pathways (higher-order features) to identify co-functioning molecular mechanisms. As called out by the MIT Tech Review article titled "Too many AI researchers think real-world problems are not relevant", there are barriers that both AI/ML and medical communities need to overcome to inform each other. This article demonstrated the feasibility to address the above needs and overcome cross-community barriers with our practical considerations of design and implementation choices by PANTHER to advance modern genetic medicine study. Interpretable and scalable ML for healthcare is just going to become more and more important. NIH dbGaP has curated many genetic medicine datasets including MESA dataset for cardiovascular diseases, whose analysis may benefit from the adoption of PANTHER. As ongoing NIH programs such as All of Us and TopMed progress towards collecting nationwide genetic medicine datasets, we expect extensive follow-up work by both communities to advance and better healthcare data analysis and learning.

\bibliographystyle{aaai21}
\bibliography{aaai21}

\end{document}